%% file: main_genRamsey.tex
\documentclass[pra,letterpaper,twocolumn,aps,footinbib,superscriptaddress,showpacs,babel]{revtex4-2}
\usepackage{physics}
\usepackage{amsmath, upgreek, amssymb, graphicx, paralist, dsfont, transparent}
\usepackage[colorlinks, linkcolor=blue, citecolor=blue, urlcolor=blue, breaklinks=true]{hyperref}
\usepackage{xspace}
\usepackage{hhline}

\usepackage{ragged2e}

\usepackage{float}
\usepackage{caption}
\usepackage[justification=justified]{subcaption}
\usepackage{overpic}

\DeclareCaptionFormat{figur}{%
  \begin{justify}
  \textbf{#1#2}#3
  \end{justify}}
\usepackage{xcolor}

\usepackage{nccmath}

\usepackage{comment}

\usepackage{enumitem}
\setlist[itemize]{leftmargin=*}

\newcommand{\nMeas}{\boldsymbol{m}}

\newcommand{\myvec}[1]{\boldsymbol{#1}}

\newcommand{\OATone}{\ensuremath{\mu_1}}
\newcommand{\OATtwo}{\ensuremath{\mu_2}}

\begin{document}

\title{Optimal Ramsey interferometry with echo protocols based on one-axis twisting}

\author{M. S. Scharnagl}
\affiliation{Institute for Theoretical Physics, Leibniz University Hannover, Appelstrasse 2, 30167 Hannover, Germany}

\author{T. Kielinski}
\affiliation{Institute for Theoretical Physics and Institute for Gravitational Physics (Albert-Einstein-Institute), Leibniz University Hannover, Appelstrasse 2, 30167 Hannover, Germany}

\author{K. Hammerer}
\affiliation{Institute for Theoretical Physics and Institute for Gravitational Physics (Albert-Einstein-Institute), Leibniz University Hannover, Appelstrasse 2, 30167 Hannover, Germany}

\begin{abstract}
We study a variational class of generalised Ramsey protocols that include two one-axis twisting (OAT) operations, one performed before the phase imprint and the other after. In this framework, we optimise the axes of the signal imprint, the OAT interactions, and the direction of the final projective measurement. We distinguish between protocols that exhibit symmetric or antisymmetric dependencies of the spin projection signal on the measured phase. Our results show that the quantum Fisher information, which sets the limits on the sensitivity achievable with a given uniaxially twisted input state, can be saturated within our class of variational protocols for almost all initial twist strengths. By incorporating numerous protocols previously documented in the literature, our approach creates a unified framework for Ramsey echo protocols with OAT states and measurements.
\end{abstract}

\date\today

\maketitle

\section{Introduction}

Quantum metrology employs quantum strategies, as e.g. entanglement and squeezing, to enhance the precision of measurements beyond classical bounds~\cite{QFI2} and has a wide range of applications, e.g. in gravitational wave detection, quantum phase estimation, quantum magnetometer, quantum spectroscopy and atomic clock synchronization \cite{qm_review}. Here we consider Ramsey interferometry as the most common method in quantum metrology with a variety of applications such as atom interferometers and optical atomic clocks. These in turn pave the way for the search for new physics, such as experiments on Lorentz violation \cite{Dreissen2022}, the search for dark matter \cite{Derevianko2014} and for variation of the fundamental constants \cite{Safronova2019}, geodesy \cite{Grotti2018} and tests of general relativity \cite{Takamoto2020}.

The precision in phase estimation achievable in a Ramsey protocol is restricted by quantum projection noise, i.e. unavoidable quantum fluctuations in a measurement. The standard Ramsey protocol using classical states is limited by the standard quantum limit (SQL). Nevertheless, it is possible to overcome this limitation up to the Heisenberg limit (HL) by using entangled or spin squeezed states, as pointed out by \citeauthor{Wineland1994}. One promising method creating spin squeezed states is one-axis twisting (OAT) \cite{Kitagawa_squeezed_1993} which can be realized experimentally through collisions in Bose-Einstein condensates \cite{Esteve2008, Gross2010}, via cavity feedback squeezing of cold atoms \cite{cavity_OAT1, cavity_OAT2} or by implementing M{\o}lmer-S{\o}rensen gates on trapped ions \cite{Blatt2008}. Besides the simple squeezing protocols which already allow to reduce the phase estimation error by a factor of $\mathcal{O}(N^{1/3})$ \cite{Kitagawa_squeezed_1993}, there have been several previous investigations on so-called echo protocols, where OAT~\cite{Leibfried2004, Leibfried2005, Davis2016, Froewis2016, Macri2016, Nolan2017, Schulte_ramsey_2020, Li2022, Volkoff2022, Colombo2022} or other squeezing methods~\cite{Burd2019, Anders2018} applied before and after the phase imprint help improving the sensitivity of the Ramsey protocol even further. 

In previous work~\cite{Schulte_ramsey_2020} we considered a variational class of echo protocols which was defined such as to allow for an analytical optimization of geometric control parameters corresponding to rotation axes and angles. Within this variational class, many of the protocols known in the literature, as well as some new protocols, could be identified as local maxima of the achievable sensitivity. This allowed a systematization of echo protocols, which, however, remained partial due to certain constraints of the variational class adopted in~\cite{Schulte_ramsey_2020}. This concerns, on the one hand, restrictions on geometric control, which excludes, for example, some of the protocols of~\cite{Froewis2016,Nolan2017}. On the other hand, the variational class was constrained to protocols whose signal $S(\phi)$ is anti-symmetric with respect to the inversion of the metrological phase, i.e. $S(-\phi)=-S(\phi)$. This constraint excluded e.g. the schemes of~\cite{Leibfried2004,Li2022} generating a symmetric signals $S(-\phi)=S(\phi)$.

Building on the protocols considered in~\cite{Schulte_ramsey_2020}, in this paper we aim at a much more general systematization of echo protocols, which is broader in terms of both their geometric degrees of freedom and the (anti)symmetry of the signal. To this purpose, we define an enlarged variational class, still based on one OAT operation each before and after signal imprint, covering all protocols studied in \cite{Leibfried2004, Leibfried2005, Davis2016, Froewis2016, Macri2016, Nolan2017, Schulte_ramsey_2020, Li2022, Volkoff2022, Colombo2022}. This generality comes at the cost of a largely numerical optimization over the variational class considered here. Our main findings are: (i) In this generalized class of Ramsey protocols, the quantum Fisher information (QFI), which bounds the maximum possible sensitivity, can be saturated for almost all initial twisting strengths $\mu\in[0,\pi]$. Here, OAT operations are described by unitaries $\mathcal{T}_{\myvec{z}}(\mu)=\exp(-i\mu S_z^2/2)$ with collective spin operator $S_z$. Saturation of the QFI is achieved by means of suitably one-axis twisted projective spin measurements. (ii) Protocols with anti-symmetric signal saturate the QFI for all twisting strengths, except in a neighborhood around $\mu\simeq \pi$. (iii) For an initial OAT around $\mu\simeq \pi$, generating GHZ-like states, the QFI is saturated by schemes with symmetric signals, with the protocols of~\cite{Leibfried2004} included as a special case.

\begin{figure*}[t]
    \centering
    \includegraphics[width=\textwidth]{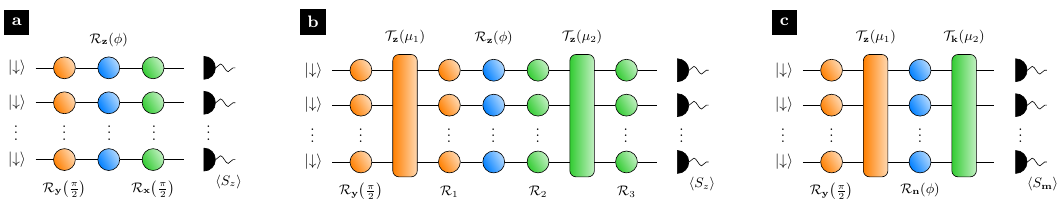}
    \vspace*{-1.1cm}
    \caption{Generalized Ramsey protocols with OTA operations $\mathcal{T}_{\myvec{z}}(\mu)=\exp(-i\mu S_z^2/2)$. Rotations $\mathcal{R}$ about an axis $\myvec{n}$ by an angle $\phi$ are denoted by $\mathcal{R}_{\myvec{n}} (\theta)$. (a) Standard Ramsey protocol without OAT. (b) Variational class of  Ramsey echo protocols with arbitrary rotations $\mathcal{R}_{1,2,3}$ and OAT strengths $\mu_1$ and $\mu_2$. (c) The same variational class, re-parameterized in terms of the axis $\myvec{n}$ of signal imprint, axis $\myvec{k}$ of second OAT, and direction $\myvec{m}$ of spin projection measurement. Here $\mathcal{R}_{1,2,3}$ do not require rotations around the z-axis.}
    \vspace*{-0.7cm}
	\label{fig:overview}
\end{figure*}

Here, we deliberately restrict our investigations to relatively simple protocols with only two squeezing operations, corresponding to the schemes demonstrated experimentally in \cite{Leibfried2005, Linnemann2016, Colombo2022}. This is complementary to the investigations in~\cite{Marciniak2022, Raphael, Thurtell2023}, which consider variational classes comprising a larger number of OAT operations before and after signal imprint. However, \cite{Marciniak2022, Raphael} investigate a reduced set of geometrical control parameters as compared to the protocols studied here. Optimization is performed here with respect to the signal-to-noise ratio achieved locally at $\phi=0$, but we do discuss the dynamic range of these optimized protocols via a figure of merit introduced in~\cite{Raphael,Leroux2017} as the effective measurement variance. The problem of Bayesian phase estimation for a given prior~\cite{Macieszczak2014} has been studied for echo-protocols in~\cite{Raphael, Thurtell2023}.

The article is organized as follows: In Sec.~\ref{sec:varclass} we introduce our general framework and describe the way in which we have generalized the Ramsey protocol. Building on that, in Sec.~\ref{sec:results} we present the local figure of merit we use for our optimization and discuss the resulting optimal protocols, including a comparison with the QFI. This reveals that the QFI can be saturated by a generalized Ramsey protocol from our variational class for almost all initial twisting strengths. To assess the experimental practicability of the optimal protocols encountered, we examine the effects of several noise sources and imperfections on their stability in Sec.~\ref{sec:noise}. In this context, we elaborate the effect of particle number fluctuations in Sec.~\ref{sec:numflucts}, the dynamic range of the optimal protocols in Sec.~\ref{sec:dynrange}, and the influence of dephasing during the OAT process in Sec.~\ref{sec:deph}. Finally, Sec.~\ref{sec:conclusion} contains a summary and an outlook on future perspectives.

\section{Variational class of interferometer protocols} \label{sec:varclass}

The variational class of protocols considered here is based on conventional Ramsey interferometry, sketched in Fig.~\ref{fig:overview}a. In this context, the dynamics of the system can be understood as the dynamics of a (pseudo) spin $\myvec{S}$ with $[S_i, S_j] = i \epsilon_{i j k} S_k$, where $i, j, k \in \{x,y,z\}$ and $\hbar = 1$. This could be an ensemble of $N$ two-level atoms, where $S_i = \frac{1}{2} \sum_{\alpha=1}^N \sigma_i^{(\alpha)}$ and $\sigma_{x,y,z}$ denote the Pauli matrices as well as the index $j$ corresponds to the $j$-th atom, but also an atomic interferometer with the modes of motion $a$ and $b$, where $S_z = a^\dagger a - b^\dagger b$, or the like. We denote rotations $\mathcal{R}_{\myvec{n}} (\theta) = e^{-i \theta S_{\myvec{n}}}$ of the total spin vector about arbitrary directions $\myvec{n} = n_x \myvec{x} + n_y \myvec{y} + n_z \myvec{z}$ with $\abs{\myvec{n}} = 1$ and angles $\theta$, where $S_{\myvec{n}} = \myvec{n} \cdot \myvec{S}$.

\begin{table*}[t]
    \centering
    \begin{tabular}{|c||c|c|c|c|c|c|c|}
        \hline
        & $\myvec{n}$ & $\myvec{m}$ & $\myvec{k}$ & $\OATone$ & $\OATtwo$ & signal symmetry & theor./expt.\\
        \hhline{|=#=|=|=|=|=|=|=|}
        \citeauthor{Kitagawa_squeezed_1993}\,\cite{Kitagawa_squeezed_1993} & $S^2$ & $S^2$ & / & $\,$ $[0, \pi]$ $\,$ & $0$ & undetermined & $\,$ T $\,$ \\
        \citeauthor{Leibfried2004}\,\cite{Leibfried2004} & -$\myvec{x}$ or $\myvec{y}$ & -$\myvec{x}$ & $\myvec{z}$ & $\pi$ & $\pi$ & symmetric & E \\
        \citeauthor{Davis2016}\,\cite{Davis2016} & $\myvec{y}$ & $\myvec{y}$ & $\myvec{z}$ & $[0, \pi]$ & $-\OATone$ & anti-symmetric & T \\
        \citeauthor{Froewis2016}\,\cite{Froewis2016} & $\myvec{y}$-$\myvec{z}$-plane & $S^2$ & $\,$ $\myvec{y}$-$\myvec{z}$-plane $\,$ & $[0, \pi]$ & $[-\pi, \pi]$ & undetermined & T \\
        \citeauthor{Macri2016}\,\cite{Macri2016} & $\myvec{y}$ & $\myvec{z}$ & $\myvec{z}$ & $[0, \pi]$ & $[0, \pi]$ & anti-symmetric & T \\
        \citeauthor{Nolan2017}\,\cite{Nolan2017} & $\,$ $\myvec{y}$-$\myvec{z}$-plane $\,$ & $\,$ $\myvec{x}$ $\,$ & $\myvec{y}$-$\myvec{z}$-plane & $[0, \pi]$ & $\,$ $[-\pi, \pi]$ $\,$ & symmetric & T \\
        \citeauthor{Schulte_ramsey_2020}\,\cite{Schulte_ramsey_2020} & $S^2$ & $S^2$ & $\myvec{z}$ & $[0, \pi]$ & $[-\pi, \pi]$ & anti-symmetric & T \\
        \citeauthor{Li2022}\,\cite{Li2022} & $\myvec{y}$ & $\myvec{x}$ & $\myvec{z}$ & $[0, \pi]$ & $-\OATone$ & symmetric & T \\
        \citeauthor{Volkoff2022}\,\cite{Volkoff2022} & $\myvec{y}$ & $\myvec{y}$ & $\myvec{z}$ & $[-\pi, 0]$ & $[-\pi, \pi]$ & anti-symmetric & T \\ 
        \citeauthor{Colombo2022}\,\cite{Colombo2022} & $\myvec{y}$ & $\myvec{y}$ & $\myvec{z}$ & $[0, 0.6]$ & $-\OATone$ & anti-symmetric & E \\
        \hline
    \end{tabular}
    \vspace*{-0.5cm}
    \caption{Echo protocols reported in the literature as characterized by their geometry (axis $\myvec{n}$ of signal imprint, axis $\myvec{k}$ of second OAT, direction $\myvec{m}$ of spin projection, cf. Fig.~\ref{fig:overview}c) and range of twisting strengths $\mu_1$ and $\mu_2$. Here, the notation $S^2$ indicates that the direction of the corresponding vector is not constraint in any way. This means it is a general 3-dimensional normalized vector, i.e. a vector in the $S^2$ sphere. The final two columns categorize the signal symmetry and indicate whether the work is theoretical or experimental.}
    \vspace*{-0.5cm}
    \label{tab:echo_references}
\end{table*}

Before introducing the generalized Ramsey protocols which we have studied in this paper, we reconsider conventional Ramsey interferometry, outlined in Fig.~\ref{fig:overview}a. This proceeds in three steps, namely state preparation (i), in which the state $\ket{\psi_{\mathrm{in}}} =\mathcal{R}_{\myvec{y}}\left(\frac{\pi}{2}\right) \ket{\downarrow}^{\otimes N} = \bigotimes_{j=1}^N \frac{\ket{\downarrow}_j + \ket{\uparrow}_j}{\sqrt{2}}$ is prepared by applying a $\frac{\pi}{2}$-pulse on the initial state $\ket{\downarrow}^{\otimes N}$, corresponding to an ensemble with all atoms in the ground state $\ket{\downarrow}$, signal imprint (ii), where the relative phase $\phi$ is imprinted on the state during the free evolution time, and measurement (iii), which consists of a second $\frac{\pi}{2}$-pulse and a measurement of $S_z$, giving an average signal of 
\begin{equation*}
    \langle S_z^{\textrm{out}}(\phi) \rangle = \bra{\psi_{\mathrm{out}}(\phi)} S_z \ket{\psi_{\mathrm{out}}(\phi)},
\end{equation*}
where $\ket{\psi_{\mathrm{out}}(\phi)} = \mathcal{R}_{\myvec{x}}\left(\frac{\pi}{2}\right) \mathcal{R}_{\myvec{z}}(\phi) \ket{\psi_{\mathrm{in}}}$ is the final state of this interferometric sequence. The challenge is to estimate the phase $\phi$ imprinted in the unitary dynamics described by $\mathcal{R}_{\myvec{z}}(\phi)$. Around the working point $\phi = 0$, the quantum projection (QPN) of the measurement is
\begin{equation*}
    (\Delta S_z^{\textrm{out}}(\phi))^2 = \bra{\psi_{\mathrm{out}}(\phi)} S_z^2 \ket{\psi_{\mathrm{out}}(\phi)} - \langle S_z^{\textrm{out}}(\phi) \rangle.
\end{equation*}
The phase estimation error can be classified by the mean squared error 
\begin{equation}\label{eq:epsilon}
    \epsilon_{\textrm{M}}(\phi) = \sum_m \left[ \Hat{\phi}(m) - \phi \right]^2 p(m \vert \phi),
\end{equation}
where $\Hat{\phi}(m)$ is the phase estimate corresponding to the measurement outcome $m$, $\phi$ the actual phase and $p(m \vert \phi)$ the conditional probability for the measurement outcome $m$ given the phase $\phi$ \cite{Raphael}. Evaluating $\epsilon_{\textrm{M}}(\phi)$ locally at the working point $\phi = 0$, using a linear phase estimator $\Hat{\phi}(m) = \frac{m}{\partial_\phi \langle S_z \rangle \vert_{\phi = 0}}$, results in (cf. Appendix \ref{app:deltaphi})
\begin{equation}\label{eq:sens}
    (\Delta \phi)^2 = \frac{(\Delta S_z)^2}{\vert \partial_{\phi} \langle S_z \rangle \vert^2}\bigg\vert_{\phi=0},
\end{equation}
which can also be obtained from Gaussian error propagation of the QPN.

While the conventional Ramsey protocol, using only uncorrelated atoms, is limited by the standard quantum limit $(\Delta \phi)^2_{\rm SQL} = 1/N$, extensions to entangled initial states can further reduce $(\Delta \phi)^2$ with the Heisenberg limit $(\Delta \phi)^2_{\rm HL} = 1/N^2$ as the fundamental lower bound. This reduction is commonly expressed in terms of the Wineland squeezing parameter \cite{Wineland1994}
\begin{equation}\label{eq:squeezingParameter}
    \xi^2 = N (\Delta \phi)^2,
\end{equation}
which takes $\xi = 1$ for conventional Ramsey interferometry.

A common method to reduce the quantum projection noise (QPN) of the standard Ramsey protocol is to perform one-axis-twisting (OAT) operations $\mathcal{T}_{\mathbf{z}}(\mu)$ during the Ramsey protocol. Here, we introduce a variational class of generalized Ramsey protocols, as shown in Fig.~\ref{fig:overview}b. As conventional Ramsey interferometry, the variational class of interferometer protocols considered here starts with (i) state preparation, consisting of rotation $\mathcal{R}_{\myvec{y}}(\pi/2)$ of the initial state $\ket{\downarrow}^{\otimes N}$ into the equatorial plane, a OAT interaction $\mathcal{T}_{\myvec{z}}(\OATone)$ with strength $\OATone$, squeezing the coherent spin state (CSS) pointing in $x$-direction, and a rotation $\mathcal{R}_1$ of the $\myvec{z}$-vector into an arbitrary direction $\myvec{n}$. This is followed by (ii) the phase imprint described by a rotation $\mathcal{R}_{\myvec{z}} (\phi)$. Finally, in (iii) a OAT measurement is performed with another rotation $\mathcal{R}_2$ turning the $\myvec{z}$-vector into a direction $\myvec{k}$, followed by a second OAT interaction $\mathcal{T}_{\myvec{z}}(\OATtwo)$ with strength $\OATtwo$ and a third rotation $\mathcal{R}_3$ turning the $\myvec{z}$-vector in a direction $\myvec{m}$. Finally, the protocol is concluded by a measurement of $S_z$.

\begin{table}[t]
    \centering
    \begin{tabular}{|c||c|c|c|}
        \hline
        signal & $\myvec{n}$ & $\myvec{m}$ & $\myvec{k}$ \\
        \hhline{|=#=|=|=|}
        $\hspace{0.05mm}$ anti-symmetric $\hspace{0.05mm}$ & $\hspace{0.05mm}$ $\myvec{y}$-$\myvec{z}$-plane $\hspace{0.05mm}$ & $\hspace{0.05mm}$ $\myvec{y}$-$\myvec{z}$-plane $\hspace{0.05mm}$ & $\hspace{0.05mm}$ $\myvec{x}$ or $\myvec{y}$-$\myvec{z}$-plane $\hspace{0.05mm}$ \\
        symmetric & $\myvec{y}$-$\myvec{z}$-plane & $\myvec{x}$ & $\myvec{x}$ or $\myvec{y}$-$\myvec{z}$-plane \\
        zero & $\myvec{x}$ & $\myvec{y}$-$\myvec{z}$-plane & $\myvec{x}$ or $\myvec{y}$-$\myvec{z}$-plane \\
        constant & $\myvec{x}$ & $\myvec{x}$ & $\myvec{x}$ \\
        no insight & $\myvec{x}$ & $\myvec{x}$ & $\myvec{y}$-$\myvec{z}$-plane \\
        \hline
    \end{tabular}
    \vspace*{-0.5cm}
    \caption{Shapes of signal for given geometrical constraints on $\myvec{n}$, $\myvec{m}$ and $\myvec{k}$, cf. Fig.~\ref{fig:overview}c.}
    \vspace*{-0.5cm}
    \label{tab:geom_constraints}
\end{table}

Choosing $\mathcal{R}_2 = \mathcal{R}_1^\dagger \Tilde{\mathcal{R}}_2$ and $\mathcal{R}_3 = \Tilde{\mathcal{R}}_2^\dagger \Tilde{\mathcal{R}}_3$, this corresponds effectively to the interferometer sequence in Fig.~\ref{fig:overview}c, which provides a more compact formal treatment, where first the state $\ket{\psi_\mathrm{in}}= \mathcal{T}_{\myvec{z}}(\OATone) \mathcal{R}_{\myvec{y}}(\pi/2)\ket{\downarrow}^{\otimes N}$ is prepared in (i) through a $\frac{\pi}{2}$-pulse operated on the initial state $\ket{\downarrow}^{\otimes N}$ followed by an OAT interaction $\mathcal{T}_{\myvec{z}}(\OATone)$ with strength $\OATone$. The signal imprint (ii) is effectively represented by a rotation $\mathcal{R}_{\myvec{n}} (\phi)$ around the axis $\myvec{n}$. After that follows the measurement phase (iii) with a second OAT interaction $\mathcal{T}_{\myvec{k}}(\OATtwo)$ with strength $\OATtwo$ creating the output state, given by
\begin{equation*}
    \ket{\psi_{\mathrm{out}}(\phi)} = \mathcal{T}_{\myvec{k}}(\OATtwo) \, \mathcal{R}_{\myvec{n}}(\phi) \, \mathcal{T}_{\mathbf{z}}(\OATone) \,  \ket{\psi_{\mathrm{in}}},
\end{equation*}
and a  measurement of $S_{\myvec{m}}$ on this state, resulting in an average signal of $\langle S_{\myvec{m}}(\phi) \rangle$
with variance $(\Delta S_{\myvec{m}})^2 (\phi)$. Consequently, this variational class of protocols depends on two twisting strengths $\OATone$ and $\OATtwo$ and three directions $\myvec{n}$, $\myvec{k}$ and $\myvec{m}$. It generalizes the standard Ramsey protocol in Fig.~\ref{fig:overview}a, and reduces to a variety of protocols discussed in the literature \cite{Kitagawa_squeezed_1993, Davis2016, Leibfried2004, Li2022, Nolan2017, Froewis2016, Macri2016, Schulte_ramsey_2020, Volkoff2022} when certain restrictions are made concerning geometry or twisting strengths, cf. Table \ref{tab:echo_references}. 

\begin{figure*}[t]
    \centering
    \vspace*{-0.3cm}
    \includegraphics[width=\textwidth]{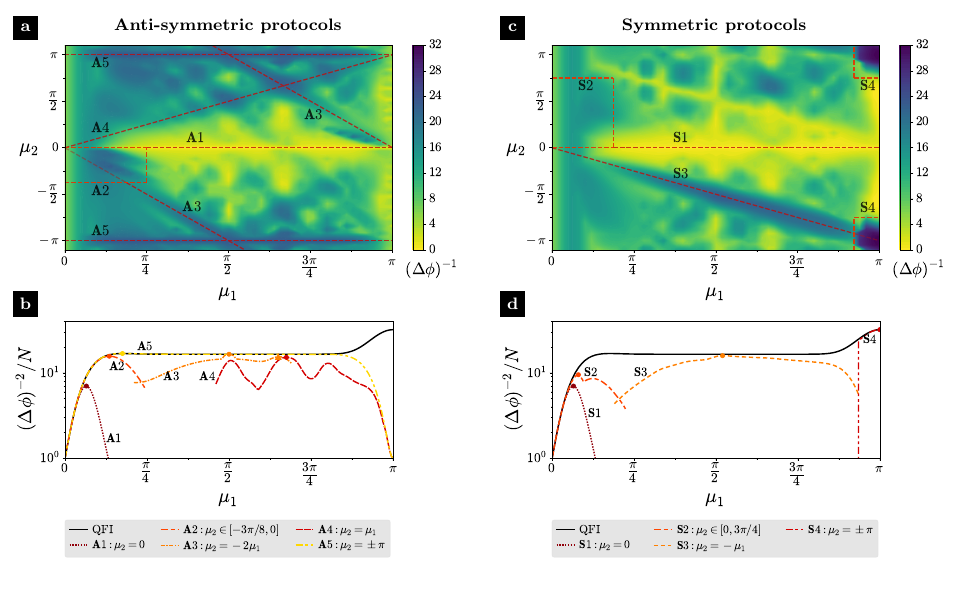}
    \vspace*{-1.3cm}
    \caption{Performance of the fully optimized anti-symmetric (a,b) and symmetric (c,d) echo protocols for $N=32$ atoms. (a,c) Phase sensitivity as quantified by the inverse measurement error $\Delta\phi^{-1}$ for given OAT strengths $\mu_1$ and $\mu_2$, optimized with respect to $\myvec{n}$, $\myvec{m}$ and $\myvec{k}$. Dashed red lines define notable anti-symmetric protocols (A1-A5) and symmetric protocols $S1-S4$. (b,d) Optimal sensitivity for protocols A1-A5 and S1-S4 referenced to the quantum Fisher information for a given initial OAT strength $\mu_1$. Markers denote protocols of maximum sensitivity whose signal shape is shown in Fig.~\ref{fig:exp_curves}. }
    \vspace*{-0.6cm}
    \label{fig:landscapesandlines}
\end{figure*}

While the protocols investigated by \cite{Schulte_ramsey_2020, Davis2016, Macri2016, Volkoff2022, Colombo2022} in general have an anti-symmetric signal curve $\langle S_{\myvec{m}}(\phi) \rangle$, the signal curves of the protocols discussed by \cite{Leibfried2004, Nolan2017, Li2022} are symmetric. We refer to protocols as being anti-symmetric or symmetric if $\langle S_{\myvec{m}}(\phi) \rangle = - \langle S_{\myvec{m}}(-\phi) \rangle$ or $\langle S_{\myvec{m}}(\phi) \rangle = \langle S_{\myvec{m}}(-\phi) \rangle$ is satisfied for all phases $\phi$, respectively. Our variational class of protocols additionally encompasses protocols without a definite signal symmetry or even with a constant signal curve. Such protocols do not produce a useful error signal, and have to be excluded from the variational class by imposing suitable conditions. For this reason, we restrict the optimization of our variational class to protocols with symmetric or anti-symmetric signal $\langle S_{\myvec{m}}(\phi) \rangle$. Following \cite{Raphael}, we find that anti-symmetry or symmetry in the signal $\langle S_{\myvec{m}}(\phi) \rangle$ can be ensured by restricting $\myvec{n}$, $\myvec{m}$ and $\myvec{k}$ to certain directions as summarized in Tab. \ref{tab:geom_constraints} (see also Appendix \ref{app:geom_const}). Only for protocols with $\myvec{n} = \myvec{m} = \myvec{x}$ and $\myvec{k}$ in the $\myvec{y}$-$\myvec{z}$-plane we gain no analytical insight on the symmetries of the underlying signals. In this case, we have to filter for anti-symmetric or symmetric protocols respectively by numerically minimizing the cosine or sine Fourier coefficients of the underlying signal curve. We note that these considerations provide sufficient (not necessary) conditions for symmetry or anti-symmetry of the signal.

\section{Optimal protocols} \label{sec:results}

\subsection{Figures of merit}

In order to optimize the variational class of protocols with anti-symmetric or symmetric signals that have been identified, we need to suitably adapt the figure of merit based on the phase measurement error of the anti-symmetric standard Ramsey protocol in Eq.~(\ref{eq:sens}). For anti-symmetric Ramsey protocols this is straight forwardly achieved for a working point $\phi=0$ by
\begin{align}\label{eq:phasevar}
    \Delta \phi(\OATone, \OATtwo,\myvec{n}, \myvec{m}, \myvec{k}) = \left.\frac{\Delta{S_{\nMeas}}}{\vert \partial_{\phi} \langle S_{\nMeas} \rangle \vert}\right|_{\phi=0}.
\end{align}
Since we want to reduce $\Delta \phi$ as far as possible, we maximize the inverse of the phase deviation $\Delta \phi$, the sensitivity 
\begin{align}
    (\Delta \phi)^{-1}_{\rm opt} (\OATone, \OATtwo) = \max_{\myvec{n}, \myvec{m}, \myvec{k}} \frac{1}{\Delta \phi(\OATone, \OATtwo,\myvec{n}, \myvec{m}, \myvec{k})},
\end{align}
with respect to the directions $\myvec{n}$, $\myvec{m}$, and $\myvec{k}$. Analogous to \cite{Schulte_ramsey_2020}, we performed the optimization over $\myvec{n}, \myvec{m}$ via a singular value decomposition (cf. Appendix \ref{app:opt}), while we executed the optimization of $\myvec{k}$ using differential evolution (DE), a numerical routine for global optimization of constrained parameters. 

\begin{figure*}
    \centering
    \vspace*{-0.3cm}
    \includegraphics[width=\textwidth]{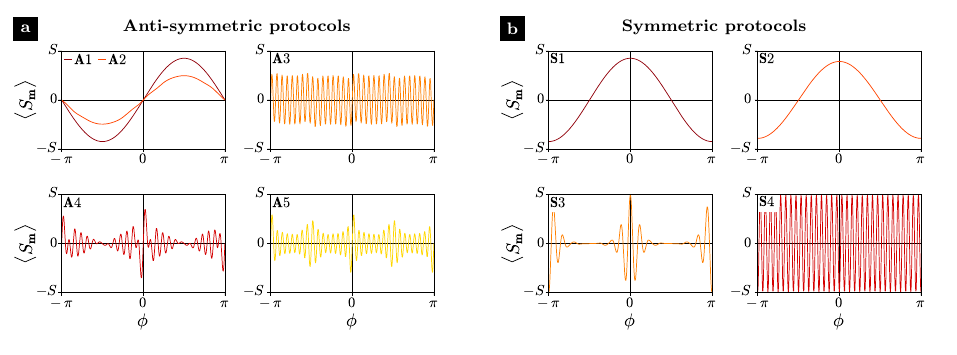}
    \vspace*{-1.2cm}
    \caption{Signal shapes $\langle S_{\myvec{m}}(\phi) \rangle$ for protocols (a) $A1-A5$ and (b) $S1-S4$ at the points of maximal sensitivity marked in Fig.~\ref{fig:landscapesandlines}b,d.}
    \vspace*{-0.5cm}
    \label{fig:exp_curves}
\end{figure*}

For symmetric Ramsey protocols the signal $\langle S_{\nMeas}(\phi) \rangle$ has an extremum at $\phi = 0$, such that the slope vanishes there. Therefore, the phase variance at $\phi=0$, as defined in Eq.~\eqref{eq:phasevar}, is no longer a meaningful measure to characterize the quality of symmetric protocols. Instead, symmetric protocols are operated with a two-point-sampling method~\cite{Leibfried2004}. This means that, in contrast to the one-point-sampling used for anti-symmetric protocols, for symmetric protocols the system is no longer probed at only one phase value $\phi = 0$, but at two points with additional phase shift $\pm \varphi$. The combined error signal after these two measurement cycles is then used to estimate the imprinted phase $\phi$. We therefore optimize the phase deviation at $\phi=\varphi \neq 0$, i.e.
\begin{align}
    \Delta \phi(\OATone, \OATtwo,\myvec{n}, \myvec{m}, \myvec{k}) = \left.\frac{\Delta{S_{\nMeas}}}{\vert \partial_{\phi} \langle S_{\nMeas} \rangle \vert}\right|_{\phi=\varphi},
\end{align}
whereby the operating point $\varphi$, as proposed by \cite{Leibfried2004}, is optimally chosen as the inflection point of the signal curve $\langle S_{\nMeas}(\phi) \rangle$, since the slope $\vert \partial_{\phi} \langle S_{\nMeas} \rangle \vert$ becomes maximum there.

\subsection{Optimal anti-symmetric protocols}   

First, we optimize the sensitivity $(\Delta \phi)^{-1}$ for the anti-symmetric protocols. In doing so, we optimize the axes $\myvec{n}$, $\myvec{m}$ and $\myvec{k}$ for given squeezing strengths $\OATone$ and $\OATtwo$ and $N = 32$ particles such that $(\Delta \phi)^{-1}$ becomes maximal. The results of this optimization are shown in Fig.~\ref{fig:landscapesandlines}a as a contour plot in the $\OATone$-$\OATtwo$-plane. In this landscape, we observe a large number of local maxima, a selection of which we refer to as protocols (A1-A5), as defined in Fig.~\ref{fig:landscapesandlines}a. Some of these local maxima (A1-A3) correspond to previously studied echo protocols known from the literature. The protocols with only initial twisting (A1), i.e. $\OATtwo=0$, correspond to the squeezing protocols discussed in \cite{Wineland1994}, while region A2 denotes protocols with low initial squeezing and small unsqueezing, and comprise the echo protocols introduced in~\cite{Davis2016}. In addition, the $\OATone=-2\OATtwo$ protocols (A3), denoting an initial twisting and a final double untwisting, comprise the anti-symmetric OUT protocols studied in \cite{Schulte_ramsey_2020}. Besides this, line A4 denotes so-called pseudo-echo protocols \cite{Nolan2017}, which do not need squeezing inversion, and the protocols along line A5 have an arbitrary initial twisting and an untwisting of strength $\OATtwo = \pm \pi$, corresponding to a projective measurement of maximally twisted Dicke states.

To better assess the magnitude of improvement in these regions, we show in Fig.~\ref{fig:landscapesandlines}b the resulting sensitivities, when optimizing over $\OATtwo$, $\myvec{n}$, $\myvec{m}$ and $\myvec{k}$ for given values of initial twisting $\OATone$. We also compare to the quantum Fisher information, which bounds the sensitivity achievable with the state $\mathcal{R}_{\myvec{n}} (\phi) \, \mathcal{T}_{\myvec{z}}(\OATone) \, \ket{\psi_{\mathrm{in}}}$, optimized over all possible rotation directions $\myvec{n}$, due to the quantum Cramér-Rao bound \cite{QFI1, QFI2}. We find that the sensitivity along line A5 saturates the QFI for all initial squeezing strengths $\OATone$ from zero initial twisting to the end of the plateau of the QFI until it slowly decreases for $\OATone$ approaching $\pi$. However, already the sensitivity of the protocols with small initial and final squeezing strength $\OATone$ and $\OATtwo$ (A2) is strongly increased compared to the simple squeezing protocols (A1). The sensitivity of region A2 saturates the QFI for every small initial squeezing strength $\OATone$ until it almost reaches the plateau value of the QFI, but starts to decrease shortly before this point. Overall, for small $\OATone$, the protocols based on twisting inversion perform significantly better than those restricted to $\OATtwo>0$. The other curves (A3, A4) approach the plateau of the QFI only for certain larger values of initial twisting $\OATone$.

\begin{figure*}[t]
    \centering
    \includegraphics[width=\textwidth]{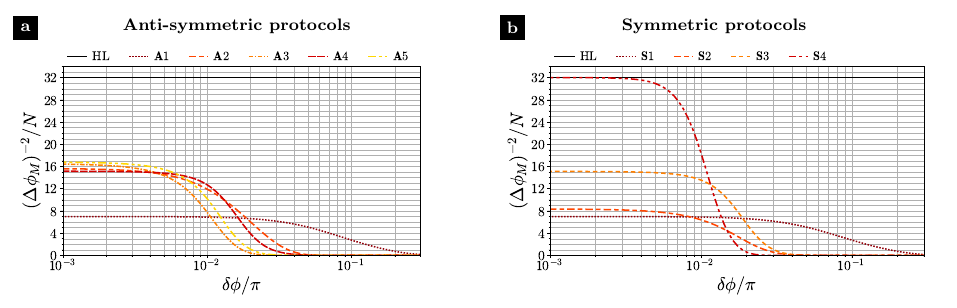}
    \vspace*{-0.7cm}
    \caption{Effective measurement variance of protocols (a) $A1-A5$ and (b) $S1-S4$ at the points of maximal sensitivity marked in Fig.~\ref{fig:landscapesandlines}b,d in comparison to Heisenberg limit (black line). For this analysis, a Gaussian phase distribution with prior variance $\delta\phi$ was assumed.}
    \vspace*{-0.5cm}
    \label{fig:emv}
\end{figure*}

\subsection{Optimal symmetric protocols}  

Analogous to the analysis of the anti-symmetric protocols, we first optimize $\myvec{n}$, $\myvec{m}$, and $\myvec{k}$ for given $\OATone$ and $\OATtwo$ and $N = 32$ particles and consider the resulting $\OATone$-$\OATtwo$ landscape (see Fig. \ref{fig:landscapesandlines}c). Here we obtain fewer local sensitivity maxima compared to the anti-symmetric case, but again select different regions (S1-S4) with comparatively high sensitivity values that contain previously studied echo protocols. As for the anti-symmetric case, the protocols with only initial squeezing (S1), i.e. $\OATtwo=0$, correspond to the squeezing protocols discussed by \cite{Wineland1994}, while now region S2 denotes the protocols with small initial squeezing and over-squeezing. In addition, the protocols with $\OATone=-\OATtwo$ (S3) corresponding to an initial twist and a final untwisting include the GESP-o protocols studied by \cite{Li2022}, and the protocols in the S4 region include the symmetric GHZ protocols considered by \cite{Leibfried2004,Leibfried2005}.

Again, we compare the maximum sensitivity values when optimizing over $\OATtwo$, $\myvec{n}$, $\myvec{m}$ and $\myvec{k}$ for given values of $\OATone$ with the quantum Fisher information (QFI) of the state $\mathcal{R}_{\myvec{n}} (\phi) \, \mathcal{T}_{\myvec{z}}(\OATone) \, \ket{\psi_{\mathrm{in}}}$ (see Fig.~\ref{fig:landscapesandlines}d). In contrast to the anti-symmetric protocols, we find only little improvement in sensitivity for the symmetric protocols with small initial and final squeezing strengths $\OATone$ and $\OATtwo$ (S2) compared to the simple squeezing protocols (S1). Here, the protocols with small $\OATone$ based on twisting inversion and those constrained to $\OATtwo>0$ perform almost equally well. Line S3 saturates the QFI only for one particular initial squeezing strength $\OATone$, almost the same as for line A3, in the middle of its plateau. However, as $\OATone$ increases, the sensitivity of line S3 remains close to the QFI until it diminishes at the end of the plateau. Region S4 represents a neighborhood of the GHZ protocols discussed by \cite{Leibfried2004, Leibfried2005}, all of which saturate the QFI for $\OATone$ near $\pi$ and eventually reach saturation of the Heisenberg limit for $\OATone = \pi$.

\section{Noise and imperfections} \label{sec:noise}

The above optimizations only consider an ideal case and disregard any noise. In this section, we will consider three important types of imperfections or limitations, namely particle number fluctuations, finite dynamic range and dephasing during twisting operations.

\subsection{Particle number fluctuations} \label{sec:numflucts}

In some platforms, e.g. neutral atom traps, the number of particles may not be precisely controlled and be subject to particle fluctuations or loss. For this reason, it is essential to consider how the optimal $\OATone$-$\OATtwo$-landscapes differ for even and odd particle numbers $N$. In general, we can find optimal axes $\myvec{n}$, $\myvec{m}$, and $\myvec{k}$ for each point of the $\mu_1$-$\mu_2$-landscape, such that the landscapes for even and odd numbers of particles appear very similar. However, the optimal axes for even and odd particle numbers are truly different at many points of the landscape. Only a few of the identified optimal protocols of Fig.~\ref{fig:landscapesandlines} are stable under particle number fluctuation, i.e. have identical optimal axes $\myvec{n}$, $\myvec{m}$ and $\myvec{k}$ for even and odd particle number. This applies for the OUT protocols (A3)~\cite{Schulte_ramsey_2020}, the protocols in the last maximum of line A4, the protocols of region A2 and S2 with small initial squeezing strength $\OATone$ and the GESP-o protocols (S3)~\cite{Li2022}. In many cases, one of the optimal directions $\myvec{n}$, $\myvec{m}$ and $\myvec{k}$ for $N=32$ has to be rotated about $\frac{\pi}{2}$ to reach the optimal sensitivity value of $N=33$, but in general the change of optimal axes from even to odd particle number varies for each point of the landscape. There is no general systematic for the variation of $\myvec{n}$, $\myvec{m}$ and $\myvec{k}$ from even to odd particle number observable. Our analysis concludes that most optimal protocols identified, with the above mentioned exceptions, are limited to experiments with stable particle numbers, such as in ion traps.

\subsection{Dynamic Range} \label{sec:dynrange}

Besides the phase measurement error $\Delta \phi$, the fringe width of the resulting signal curve $\langle S_{\myvec{m}}(\phi) \rangle$ plays an important role for the applicability of the protocols in the experiment. In optical atomic clocks, for example, a small fringe width increases the probability of the occurrence of fringe hops, which then in turn limit the stability of the clock~\cite{Schulte2020}. In Fig.~\ref{fig:exp_curves}a and \ref{fig:exp_curves}b, we show the shape of the average signal $\langle S_{\myvec{m}} (\phi)\rangle$ with optimized $\OATone$, $\OATtwo$, $\myvec{n}$, $\myvec{m}$ and $\myvec{k}$ corresponding to the maxima in sensitivity marked in Fig.~\ref{fig:landscapesandlines}b and \ref{fig:landscapesandlines}d for the anti-symmetric (A1-A5) and symmetric (S1-S4) protocols. This reveals that the central fringe can become quite narrow for both anti-symmetric and symmetric protocols using large twisting strengths, which limits the dynamical range of the interferometer. 

\begin{figure*}[t]
    \centering
    \includegraphics[width=\textwidth]{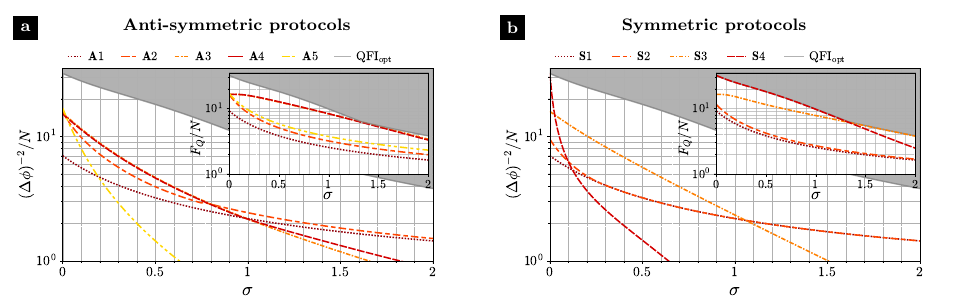}
    \vspace*{-0.7cm}
    \caption{Sensitivity with increasing dephasing strength $\sigma$ of protocols (a) $A1-A5$ and (b) $S1-S4$ at the points of maximal sensitivity marked in Fig.~\ref{fig:landscapesandlines}b,d in comparison to the QFI of a dephased input state with equal initial twisting strength $\mu_1$ (inset).}
    \vspace*{-0.5cm}
    \label{fig:deph}
\end{figure*}

To quantify the trade-off between enhancement in sensitivity and reduction of the dynamical range, we use the effective measurement variance $(\Delta \phi_M)^2$ defined by~\cite{Raphael}
\begin{equation*}
    \frac{1}{(\Delta \phi_M)^2} = \frac{1}{\epsilon_{\textrm{B}}(\phi)} - \mathcal{I}.
\end{equation*}
Here, the Bayesian mean squared error for a given prior phase distribution $\mathcal{P}(\phi)$ is
\begin{equation*}
    \epsilon_{\textrm{B}}(\phi) = \int_{-\infty}^\infty \textrm{d}\phi \, \epsilon_{\textrm{M}}(\phi) \, \mathcal{P}(\phi),
\end{equation*}
where $\epsilon_{\textrm{M}}(\phi)$ was defined in Eq.~\eqref{eq:epsilon} and $\mathcal{I}$ is the Fisher information of the prior distribution. The effective measurement variance satisfies $(\Delta \phi_M)^2 \geq \frac{1}{\, \overline{F_Q} \,}$, similar to the quantum Cramér-Rao bound, where $\overline{F_Q}$ denotes the Fisher information averaged over the prior distribution \cite{Raphael}. $\Delta \phi_M$ therefore quantifies the true phase estimation error of a single measurement tracing out the prior knowledge. In Figure \ref{fig:emv}, we show the effective measurement variance in dependence of the prior width $\delta \phi$ of a Gaussian laser phase distribution, i.e. the sensitivity of measurement protocols for increasing dynamical range. This shows that for a small prior phase variance $\delta \phi$, the protocols ${A2}$-${A5}$ and ${S2}$-${S4}$ lead to high improvements in sensitivity compared to the squeezing protocols ${A1}/{S1}$. With increasing prior phase variance $\delta \phi$, the advantage of the protocols ${A2}$-${A5}$ and ${S2}$-${S4}$ over the squeezing protocols decreases until the sensitivity of the squeezing protocols prevails due to their smaller dynamical range.

\subsection{Dephasing during twisting} \label{sec:deph}

Another significant source of noise is the dephasing that occurs during the twisting. Therefore, we consider how the sensitivity changes as dephasing increases for each of the local maxima in sensitivity corresponding to the regions A1-A5 and S1-S4. We compare the achieved sensitivity to the QFI of a dephased input state with the same initial twisting strength $\mu_1$, cf. Fig.~\ref{fig:deph}. Dephasing during the twisting process is described by the master equation 
\begin{align*}
    \frac{\partial}{\partial t} \rho &= -i [H, \rho] + \gamma \left[ L \rho L^\dagger - \frac{1}{2} L^\dagger L \rho - \frac{1}{2} \rho L^\dagger L \right],
\end{align*}
where $H = \chi S_{\myvec{a}}^2$ with $\frac{\mu}{2} = \chi t$ is the OAT Hamiltonian and $L = S_{\myvec{a}}$.  The dephasing strength is defined as dimensionless parameter $\sigma = \frac{\gamma}{\vert \chi \vert}$. Here, $\myvec{a} = \myvec{z}$ for the first and $\myvec{a} = \myvec{k}$ for the second OAT operation. 

First, we observe that with increasing dephasing, the $\mu_1$ value of the corresponding sensitivity maximum as well as the optimal axes $\myvec{n}$, $\myvec{m}$ and $\myvec{k}$ undergo small changes. Thus, we optimize $\mu_1$ and the axes $\myvec{n}$, $\myvec{m}$ and $\myvec{k}$ in a small range around the original value and direction at $\sigma = 0$, respectively. Our analysis reveals that the sensitivity maximum of A3 at $\mu_1 \approx \pi/2$ is very unstable under dephasing. Therefore, in Fig.~\ref{fig:deph}a we show the sensitivity of the second maximum of A3, which is marked with a triangle in Fig.~\ref{fig:landscapesandlines}b. However, due to the similar signal shape and dynamic range as the maximum of A4, we have not included these protocols in Fig.~\ref{fig:exp_curves},~\ref{fig:emv}. Moreover, we find that the sensitivity at the maxima of A2, A4 and S3 is relatively stable under dephasing, while the sensitivity at the maxima of A5 and S4 decreases very rapidly with increasing dephasing. At a dephasing strength of $\sigma \approx 1$, the sensitivity of all maxima, except the maximum of A2, falls below the sensitivity of the simple squeezing protocols A1, S1. Surprisingly, we find that the sensitivity of the maximum of A2 surpasses that of A1 for all dephasing strengths considered up to $\sigma = 2$. This shows that experiments with strong dephasing during the twisting process can benefit from using protocols with small twisting and untwisting, respectively. 

However, in comparison to the QFI of a dephased input state, we find that the sensitivities of regions A3-A5 and S3-S4 quickly diverge from the QFI with increasing dephasing, while the sensitivities of the simple squeezing protocols A1, S1 exhibit a constant offset to the corresponding QFI. Therefore, dephasing during the second OAT process results in high losses compared to the possible achievable sensitivity with the corresponding dephased input state.

\section{Summary and Outlook} \label{sec:conclusion}

In conclusion, we have provided a comprehensive systematization of one-axis twisting echo protocols, building upon the results of our earlier work \cite{Schulte_ramsey_2020}. We accounted for and optimized a larger set of geometrical control parameters, also considering protocols with symmetric signals. Although this approach limited us to primarily numerical optimization, we identified a larger number of local sensitivity maxima compared to \cite{Schulte_ramsey_2020}. A significant finding is that the class of entangled measurements considered in this study enables the saturation of the QFI for nearly all initial squeezing strengths $\OATone$. This was not achievable in the protocols discussed in \cite{Schulte_ramsey_2020}, emphasizing the importance of optimizing the direction, $\myvec{k}$, of the second twisting. Most of the QFI is saturated by protocols with anti-symmetric signals, except near $\OATone = \pi$, where only protocols with symmetric signals can saturate the QFI.

Furthermore, we analyzed the impact of various noise sources, including particle number fluctuations, prior phase noise, and dephasing noise during the twisting process, on the sensitivity of these optimal protocols. We discovered that the sensitivity for the majority of the identified protocols is not stable under particle number fluctuations; hence, these protocols are most suitable for experiments where the particle number, $N$, is well-controlled. For experiments with substantial particle number fluctuations, only the OUT protocols discussed in \cite{Schulte_ramsey_2020} (A3), protocols at the maximum of region A4, and protocols with minor twisting and untwisting (A2) are appropriate. Regarding dephasing noise during the twisting operation, we demonstrated that protocols in the maxima of regions A2, A3, A4, and S3 are especially stable. Interestingly, even with significant dephasing noise, the sensitivity of the optimal A2 region protocols can surpass that of the simple squeezing protocols (A1, S1). Our examination of the trade-off between sensitivity gain and dynamic range loss revealed that the protocols at the maximum of regions A2 and A4, along with the optimal protocols of region S3, exhibit the greatest resilience to prior phase noise. However, with extensive prior phase noise, only the simple squeezing protocols (A1, S1) maintain non-vanishing sensitivity. An optimization using Bayesian phase estimation in the variational class of echo-protocols discussed here, along with its application in optical atomic clocks, will be addressed in subsequent work.

\section*{Acknowledgements}

We acknowledge funding by the Deutsche Forschungsgemeinschaft (DFG, German Research Foundation) through Project-ID 274200144 – SFB 1227 (projects A06 and A07) and Project-ID 390837967 - EXC 2123, and by the Quantum Valley Lower Saxony Q1 project (QVLS-Q1) through the Volkswagen foundation and the ministry for science and culture of Lower Saxony.

\input{Refs.bbl}

\appendix
\onecolumngrid

\vspace{1cm}

\section{Optimization} \label{app:opt}

Analogous to \citeauthor{Schulte_ramsey_2020} \cite{Schulte_ramsey_2020}, we can rewrite
\begin{equation*}
    (\Delta S_{\myvec{m}})^2 = \myvec{m}^T Q \myvec{m} \quad \textrm{and} \quad \Big\vert \frac{\partial  \langle S_{\myvec{m}} \rangle}{\partial \phi} \big\vert_{\phi = 0} \Big\vert = \myvec{n}^T M \myvec{m}
\end{equation*}
with the help of the matrices $M$ and $Q$, 
where $M_{kl} = i \langle [S_k(\OATone), S_l(\OATone + \OATtwo)] \rangle \vert_{\phi=0}$ and $Q_{kl} = \langle S_k(\OATone + \OATtwo) S_l(\OATone + \OATtwo) - \langle S_k(\OATone + \OATtwo) \rangle \langle S_l(\OATone + \OATtwo) \rangle \rangle \vert_{\phi = 0}$, and optimise 
\begin{equation*}
    (\Delta \phi)^{-1} = \frac{\myvec{n}^T M \myvec{m}}{\sqrt{\myvec{m}^T Q \myvec{m}}} = \frac{\myvec{n}^T M Q^{-1/2} \myvec{v}}{\sqrt{\myvec{v}^T \myvec{v}}} = \myvec{n}^T M Q^{-1/2} \frac{\myvec{v}}{\Vert \myvec{v} \Vert} = \myvec{n}^T M Q^{-1/2} \myvec{e}
\end{equation*}
via a singular value decomposition (SVD) of the matrix $M Q^{-1/2} = U \Sigma V^\ast$. Hereby, the maximum singular value corresponds to the maximum value $(\Delta \phi)_{\textrm{max}}^{-1} = \sigma_{\textrm{max}}$ and the optimal axes $\myvec{n}$ and $\myvec{m}$ are given by 
\begin{equation*}
    \myvec{n} = \mathfrak{Re}\{\myvec{u}_{\textrm{max}}^T\} \quad \textrm{and} \quad \myvec{m} = \mathfrak{Re}\{Q^{1/2} \myvec{v}_{\textrm{max}}\}.
\end{equation*}
Due to the increased complexity of $\langle S_{\myvec{m}} \rangle$ and $(\Delta S_{\myvec{m}})^2$, we cannot evaluate the matrices $M$ and $Q$ analytically via the characteristic function $X_A = \bra{\theta, \varphi} e^{\gamma S_-} e^{\beta S_z} e^{\alpha S_+} \ket{\theta, \varphi}$ \cite{Arrechi1972}, as \citeauthor{Schulte_ramsey_2020} \cite{Schulte_ramsey_2020} did, but had to evaluate them numerically by identifying $S_x$, $S_y$ and $S_z$ with $(N+1)\times(N+1)$-matrices and the states $\ket{m}$, $m=-N/2, ..., N/2$, with $(N+1)$-dimensional vectors.

\section{Symmetry requirements} \label{app:geom_const}

\noindent Inspired by the work of \citeauthor{Raphael} \cite{Raphael}, we introduce the unitary operator $\mathcal{U} = \mathcal{P}_x = e^{-i \pi S_x}$. Since $\langle S_{\Vec{m}} \rangle (\phi)$ is given by
\begin{equation*}
    \langle S_{\Vec{m}} \rangle (\phi) = \langle N/2 \vert_x \, \mathcal{T}_{\Vec{e}_z}^\dagger(\mu_1) \, \mathcal{R}_{\Vec{n}}^{\dagger}(\phi) \, \mathcal{T}_{\Vec{k}}^{\dagger}(\mu_2) \, S_{\Vec{m}} \, \mathcal{T}_{\Vec{k}}(\mu_2) \, \mathcal{R}_{\Vec{n}}(\phi) \, \mathcal{T}_{\Vec{e}_z}(\mu_1) \, \vert N/2 \rangle_x,
\end{equation*}
we can obtain a sufficient condition for symmetry / anti-symmetry around the working point $\langle S_{\Vec{m}} \rangle(\phi = 0)$ by looking for geometrical restrictions on $\myvec{n}$, $\myvec{m}$ and $\myvec{k}$ so that 
\begin{align*}
    \mathcal{U} \, \mathcal{R}_{\Vec{n}}(\phi) \, \mathcal{U}^{\dagger} = \mathcal{R}_{\Vec{n}}(-\phi), \quad \mathcal{U} \, S_{\Vec{m}} \, \mathcal{U}^{\dagger} = \pm S_{\Vec{m}}, \\
    \mathcal{U} \, \mathcal{T}_{\Vec{e}_z}(\mu_1) \, \mathcal{U}^{\dagger} = \mathcal{T}_{\Vec{e}_z}(\mu_1), \quad 
    \mathcal{U} \, \mathcal{T}_{\Vec{k}}(\mu_2) \, \mathcal{U}^{\dagger} = \mathcal{T}_{\Vec{k}}(\mu_2)
\end{align*}
is fulfilled. Because $\vert N/2 \rangle_x$ is an eigenstate in the $S_x$-basis, we find
\begin{equation*}
    \mathcal{P}_x \, \vert N/2 \rangle_x = e^{-i \pi S_x} \, \vert N/2 \rangle_x = e^{-i \, \frac{\pi N}{2}} \, \vert N/2 \rangle_x.
\end{equation*}
With $\sigma_x \sigma_x \sigma_x = \sigma_x$, $\sigma_x \sigma_y \sigma_x = i \sigma_z \sigma_x = - \sigma_y$ and $\sigma_x \sigma_z \sigma_x = - i \sigma_y \sigma_x = - \sigma_z$ we obtain
\begin{align*}
    \mathcal{P}_x \, S_x \, \mathcal{P}_x^{\dagger} &= {\displaystyle \prod_{j, k=1}^N} e^{-i \, \frac{\pi}{2} \sigma_x^{(j)}} \, S_x \, e^{i \, \frac{\pi}{2} \sigma_x^{(k)}} = {\displaystyle \prod_{j, k=1}^N} \left( \frac{1}{2} {\displaystyle \sum_{l=1}^N} \sigma_x^{(j)} \sigma_x^{(l)} \sigma_x^{(k)} \right) = \frac{1}{2} {\displaystyle \sum_{l=1}^N} \sigma_x^{(l)} = S_x, \\
    \mathcal{P}_x \, S_{y,z} \, \mathcal{P}_x^{\dagger} &= {\displaystyle \prod_{j, k=1}^N} e^{-i \, \frac{\pi}{2} \sigma_x^{(j)}} \, S_{y,z} \, e^{i \, \frac{\pi}{2} \sigma_x^{(k)}} = {\displaystyle \prod_{j, k=1}^N} \left( \frac{1}{2} {\displaystyle \sum_{l=1}^N} \sigma_x^{(j)} \sigma_{y,z}^{(l)} \sigma_x^{(k)} \right) = - \frac{1}{2} {\displaystyle \sum_{l=1}^N} \sigma_{y,z}^{(l)} = - S_{y,z}
\end{align*}
and from that it directly follows that
\begin{align*}
    \mathcal{P}_x \, S_x^2 \, \mathcal{P}_x^{\dagger} &= \mathcal{P}_x \, S_x \, \mathcal{P}_x^{\dagger} \, \mathcal{P}_x \, S_x \, \mathcal{P}_x^{\dagger} = S_x \cdot S_x = S_x^2 \\
    \mathcal{P}_x \, S_{y,z}^2 \, \mathcal{P}_x^{\dagger} &= \mathcal{P}_x \, S_{y,z} \, \mathcal{P}_x^{\dagger} \, \mathcal{P}_x \, S_{y,z} \, \mathcal{P}_x^{\dagger} = (- S_{y,z}) \cdot (- S_{y,z}) = S_{y,z}^2.
\end{align*}
\noindent Thus, we derive
\begin{equation*}
    \mathcal{P}_x \, S_{\Vec{m}} \, \mathcal{P}_x^{\dagger} = m_x S_x - m_y S_y - m_z S_z.
\end{equation*}
So if we restrict $S_{\myvec{m}}$ to the $x$-direction ($m_x=1$, $m_y=m_z=0$), we obtain
\begin{equation*}
    \mathcal{P}_x \, S_{\Vec{m}} \, \mathcal{P}_x^{\dagger} = m_x S_x = S_{\Vec{m}},
\end{equation*}
as well as we find
\begin{equation*}
    \mathcal{P}_x \, S_{\Vec{m}} \, \mathcal{P}_x^{\dagger} = - m_y S_y - m_z S_z = - S_{\Vec{m}} 
\end{equation*}
when restricting $S_{\myvec{m}}$ to the $y$-$z$-plane ($m_x=0$, $m_y=\cos(\beta)$, $m_z=\sin(\beta)$). With this we can follow
\begin{equation*}
    \mathcal{P}_x \, \mathcal{R}_{\Vec{n}}(\phi) \, \mathcal{P}_x^{\dagger} = {\displaystyle \sum_{j=1}^{\infty}} \frac{(-i \phi)^j}{j!} \, \mathcal{P}_x \, S_{\Vec{n}}^j \, \mathcal{P}_x^{\dagger} = {\displaystyle \sum_{j=1}^{\infty}} \frac{(-i \phi)^j}{j!} \, \left( \mathcal{P}_x \, S_{\Vec{n}} \, \mathcal{P}_x^{\dagger} \right)^j = {\displaystyle \sum_{j=1}^{\infty}} \frac{(-i \phi)^j}{j!} \, \left( n_x S_x - n_y S_y - n_z S_z \right)^j.
\end{equation*}
Hence, restricting $\myvec{n}$ to the $x$-direction gives us
\begin{equation*}
    \mathcal{P}_x \, \mathcal{R}_{\Vec{n}}(\phi) \, \mathcal{P}_x^{\dagger} = {\displaystyle \sum_{j=1}^{\infty}} \frac{(-i \phi)^j}{j!} \, (n_x S_x)^j = {\displaystyle \sum_{j=1}^{\infty}} \frac{(-i \phi)^j}{j!} \, S_{\Vec{n}}^j = \mathcal{R}_{\Vec{n}}(\phi)
\end{equation*}
while a restriction of $\myvec{n}$ to the $y$-$z$-plane leads to
\begin{equation*}
    \mathcal{P}_x \, \mathcal{R}_{\Vec{n}}(\phi) \, \mathcal{P}_x^{\dagger} = {\displaystyle \sum_{j=1}^{\infty}} \frac{(-i \phi)^j}{j!} \, (- n_y S_y - n_z S_z)^j = {\displaystyle \sum_{j=1}^{\infty}} \frac{(-i \phi)^j}{j!} \, (- S_{\Vec{n}})^j = \mathcal{R}_{\Vec{n}}(-\phi).
\end{equation*}
For the direction of the second OAT, we find that 
\begin{align*}
    \mathcal{P}_x \, &\mathcal{T}_{\Vec{k}}(\mu_2) \, \mathcal{P}_x^\dagger = {\displaystyle \sum_{j=1}^{\infty}} \frac{(-i \, (\mu_2/2))^j}{j!} \mathcal{P}_x \, ((k_x S_x + k_y S_y + k_z S_z)^2)^j \, \mathcal{P}_x^{\dagger} = {\displaystyle \sum_{j=1}^{\infty}} \frac{(-i \, (\mu_2/2))^j}{j!} \, ((k_x S_x - k_y S_y - k_z S_z)^2)^j.
\end{align*}
So for both cases, that $\myvec{k}$ is restricted to the $x$-direction and $\myvec{k}$ lies in the $y$-$z$-plane, we obtain
\begin{align*}
    \mathcal{P}_x \, &\mathcal{T}_{\Vec{k}}(\mu_2) \, \mathcal{P}_x^\dagger = {\displaystyle \sum_{j=1}^{\infty}} \frac{(-i \, (\mu_2/2))^j}{j!} ((k_x S_x)^2)^j = {\displaystyle \sum_{j=1}^{\infty}} \frac{(-i \, (\mu_2/2))^j}{j!} (S_{\Vec{k}}^2)^j = \mathcal{T}_{\Vec{k}}(\mu_2)
\end{align*}
and 
\begin{align*}
    \mathcal{P}_x \, &\mathcal{T}_{\Vec{k}}(\mu_2) \, \mathcal{P}_x^\dagger = {\displaystyle \sum_{j=1}^{\infty}} \frac{(-i \, (\mu_2/2))^j}{j!} \, ((- k_y S_y - k_z S_z)^2)^j = {\displaystyle \sum_{j=1}^{\infty}} \frac{(-i \, (\mu_2/2))^j}{j!} ((-S_{\Vec{k}})^2)^j = {\displaystyle \sum_{j=1}^{\infty}} \frac{(-i \, (\mu_2/2))^j}{j!} (S_{\Vec{k}}^2)^j = \mathcal{T}_{\Vec{k}}(\mu_2)
\end{align*}
respectively.

\noindent From these calculations we infer that restricting the vectors $\Vec{n}$ to the $y$-$z$-plane and $\Vec{m}$ to the $x$-direction as well as $\Vec{k}$ to the $x$-direction or the $y$-$z$-plane is sufficient to assure symmetry around $\langle S_{\Vec{m}} \rangle (\phi=0)$, since
\begin{align*}
    \langle S_{\Vec{m}} \rangle (\phi) &= \langle N/2 \vert_x \, \mathcal{T}_{\Vec{e}_z}^{\dagger}(\mu_1) \, \mathcal{R}_{\Vec{n}}^{\dagger}(\phi) \, \mathcal{T}_{\Vec{k}}^{\dagger}(\mu_2) \, S_{\Vec{m}} \, \mathcal{T}_{\Vec{k}}(\mu_2) \, \mathcal{R}_{\Vec{n}}(\phi) \, \mathcal{T}_{\Vec{e}_z}(\mu_1) \, \vert N/2 \rangle_x \\
    &= \langle N/2 \vert_x \, \mathcal{P}_x^\dagger \, \mathcal{P}_x \, \mathcal{T}_{\Vec{e}_z}^{\dagger}(\mu_1) \, \mathcal{P}_x^\dagger \, \mathcal{P}_x \, \mathcal{R}_{\Vec{n}}^{\dagger}(\phi) \, \mathcal{P}_x^\dagger \, \mathcal{P}_x \, \mathcal{T}_{\Vec{k}}^{\dagger}(\mu_2) \, \mathcal{P}_x^\dagger \, \mathcal{P}_x \, S_{\Vec{m}} \, \mathcal{P}_x^\dagger \, \mathcal{P}_x \, \mathcal{T}_{\Vec{k}}(\mu_2) \, \mathcal{P}_x^\dagger \\
    &\qquad \qquad \, \, \mathcal{P}_x \, \mathcal{R}_{\Vec{n}}(\phi) \, \mathcal{P}_x^\dagger \, \mathcal{P}_x \, \mathcal{T}_{\Vec{e}_z}(\mu_1) \, \mathcal{P}_x^{\dagger} \, \mathcal{P}_x \, \vert N/2 \rangle_x \\
    &= \langle N/2 \vert_x \, e^{i \, \frac{\pi N}{2}} \, \mathcal{T}_{\Vec{e}_z}^{\dagger}(\mu_1) \, \mathcal{R}_{\Vec{n}}^{\dagger}(-\phi) \, \mathcal{T}_{\Vec{k}}^{\dagger}(\mu_2) \, S_{\Vec{m}} \, \mathcal{T}_{\Vec{k}}(\mu_2) \, \mathcal{R}_{\Vec{n}}(-\phi) \, \mathcal{T}_{\Vec{e}_z}(\mu_1) \, e^{-i \, \frac{\pi N}{2}} \, \vert N/2 \rangle_x \\
    &= \langle S_{\Vec{m}} \rangle (-\phi)
\end{align*}
and restricting $\Vec{n}$ and $\Vec{m}$ to the $y$-$z$-plane as well as $\Vec{k}$ to the $x$-direction or the $y$-$z$-plane is sufficient to assure anti-symmetry around $\langle S_{\Vec{m}} \rangle (\phi=0)$, since
\begin{align*}
    \langle S_{\Vec{m}} \rangle (\phi) &= \langle N/2 \vert_x \, \mathcal{T}_{\Vec{e}_z}^{\dagger}(\mu_1) \, \mathcal{R}_{\Vec{n}}^{\dagger}(\phi) \, \mathcal{T}_{\Vec{k}}^{\dagger}(\mu_2) \, S_{\Vec{m}} \, \mathcal{T}_{\Vec{k}}(\mu_2) \, \mathcal{R}_{\Vec{n}}(\phi) \, \mathcal{T}_{\Vec{e}_z}(\mu_1) \, \vert N/2 \rangle_x \\
    &= \langle N/2 \vert_x \, \mathcal{P}_x^\dagger \, \mathcal{P}_x \, \mathcal{T}_{\Vec{e}_z}^{\dagger}(\mu_1) \, \mathcal{P}_x^\dagger \, \mathcal{P}_x \, \mathcal{R}_{\Vec{n}}^{\dagger}(\phi) \, \mathcal{P}_x^\dagger \, \mathcal{P}_x \, \mathcal{T}_{\Vec{k}}^{\dagger}(\mu_2) \, \mathcal{P}_x^\dagger \, \mathcal{P}_x \, S_{\Vec{m}} \, \mathcal{P}_x^\dagger \, \mathcal{P}_x \, \mathcal{T}_{\Vec{k}}(\mu_2) \, \mathcal{P}_x^\dagger \\
    &\qquad \qquad \, \, \mathcal{P}_x \, \mathcal{R}_{\Vec{n}}(\phi) \, \mathcal{P}_x^\dagger \, \mathcal{P}_x \, \mathcal{T}_{\Vec{e}_z}(\mu_1) \, \mathcal{P}_x^{\dagger} \, \mathcal{P}_x \, \vert N/2 \rangle_x \\
    &= - \langle N/2 \vert_x \, e^{i \, \frac{\pi N}{2}} \, \mathcal{T}_{\Vec{e}_z}^{\dagger}(\mu_1) \, \mathcal{R}_{\Vec{n}}^{\dagger}(-\phi) \, \mathcal{T}_{\Vec{k}}^{\dagger}(\mu_2) \, S_{\Vec{m}} \, \mathcal{T}_{\Vec{k}}(\mu_2) \, \mathcal{R}_{\Vec{n}}(-\phi) \, \mathcal{T}_{\Vec{e}_z}(\mu_1) \, e^{-i \, \frac{\pi N}{2}} \, \vert N/2 \rangle_x \\
    &= - \langle S_{\Vec{m}} \rangle (-\phi).
\end{align*}
For $\Vec{n}$ and $\Vec{m}$ in $x$-direction as well as $\Vec{k}$ in $x$-direction or in the $y$-$z$-plane, we obtain
\begin{align*}
    \langle S_{\Vec{m}} \rangle (\phi) &= \langle N/2 \vert_x \, \mathcal{T}_{\Vec{e}_z}^{\dagger}(\mu_1) \, \mathcal{R}_{\Vec{n}}^{\dagger}(\phi) \, \mathcal{T}_{\Vec{k}}^{\dagger}(\mu_2) \, S_{\Vec{m}} \, \mathcal{T}_{\Vec{k}}(\mu_2) \, \mathcal{R}_{\Vec{n}}(\phi) \, \mathcal{T}_{\Vec{e}_z}(\mu_1) \, \vert N/2 \rangle_x \\
    &= \langle N/2 \vert_x \, \mathcal{P}_x^\dagger \, \mathcal{P}_x \, \mathcal{T}_{\Vec{e}_z}^{\dagger}(\mu_1) \, \mathcal{P}_x^\dagger \, \mathcal{P}_x \, \mathcal{R}_{\Vec{n}}^{\dagger}(\phi) \, \mathcal{P}_x^\dagger \, \mathcal{P}_x \, \mathcal{T}_{\Vec{k}}^{\dagger}(\mu_2) \, \mathcal{P}_x^\dagger \, \mathcal{P}_x \, S_{\Vec{m}} \, \mathcal{P}_x^\dagger \, \mathcal{P}_x \, \mathcal{T}_{\Vec{k}}(\mu_2) \, \mathcal{P}_x^\dagger \\
    &\qquad \qquad \, \, \mathcal{P}_x \, \mathcal{R}_{\Vec{n}}(\phi) \, \mathcal{P}_x^\dagger \, \mathcal{P}_x \, \mathcal{T}_{\Vec{e}_z}(\mu_1) \, \mathcal{P}_x^{\dagger} \, \mathcal{P}_x \, \vert N/2 \rangle_x \\
    &= \langle N/2 \vert_x \, e^{i \, \frac{\pi N}{2}} \, \mathcal{T}_{\Vec{e}_z}^{\dagger}(\mu_1) \, \mathcal{R}_{\Vec{n}}^{\dagger}(\phi) \, \mathcal{T}_{\Vec{k}}^{\dagger}(\mu_2) \, S_{\Vec{m}} \, \mathcal{T}_{\Vec{k}}(\mu_2) \, \mathcal{R}_{\Vec{n}}(\phi) \, \mathcal{T}_{\Vec{e}_z}(\mu_1) \, e^{-i \, \frac{\pi N}{2}} \, \vert N/2 \rangle_x \\
    &= \langle S_{\Vec{m}} \rangle (\phi),
\end{align*}
i.e. we do not gain any further insight. These protocols can be symmetric or anti-symmetric but must not necessarily have any symmetry properties at all. Here, we need to minimize the real or the imaginary Fourier coefficients respectively to cull the symmetric and anti-symmetric protocols of this set.

\noindent In the case, that $\myvec{n}$ is in $x$-direction, $\myvec{m}$ in the $y$-$z$-plane and $\myvec{k}$ in $x$-direction or in the $y$-$z$-plane, we obtain
\begin{align*}
    \langle S_{\Vec{m}} \rangle (\phi) &= \langle N/2 \vert_x \, \mathcal{T}_{\Vec{e}_z}^{\dagger}(\mu_1) \, \mathcal{R}_{\Vec{n}}^{\dagger}(\phi) \, \mathcal{T}_{\Vec{k}}^{\dagger}(\mu_2) \, S_{\Vec{m}} \, \mathcal{T}_{\Vec{k}}(\mu_2) \, \mathcal{R}_{\Vec{n}}(\phi) \, \mathcal{T}_{\Vec{e}_z}(\mu_1) \, \vert N/2 \rangle_x \\
    &= \langle N/2 \vert_x \, \mathcal{P}_x^\dagger \, \mathcal{P}_x \, \mathcal{T}_{\Vec{e}_z}^{\dagger}(\mu_1) \, \mathcal{P}_x^\dagger \, \mathcal{P}_x \, \mathcal{R}_{\Vec{n}}^{\dagger}(\phi) \, \mathcal{P}_x^\dagger \, \mathcal{P}_x \, \mathcal{T}_{\Vec{k}}^{\dagger}(\mu_2) \, \mathcal{P}_x^\dagger \, \mathcal{P}_x \, S_{\Vec{m}} \, \mathcal{P}_x^\dagger \, \mathcal{P}_x \, \mathcal{T}_{\Vec{k}}(\mu_2) \, \mathcal{P}_x^\dagger \\
    &\qquad \qquad \, \, \mathcal{P}_x \, \mathcal{R}_{\Vec{n}}(\phi) \, \mathcal{P}_x^\dagger \, \mathcal{P}_x \, \mathcal{T}_{\Vec{e}_z}(\mu_1) \, \mathcal{P}_x^{\dagger} \, \mathcal{P}_x \, \vert N/2 \rangle_x \\
    &= \langle N/2 \vert_x \, e^{i \, \frac{\pi N}{2}} \, \mathcal{T}_{\Vec{e}_z}^{\dagger}(\mu_1) \, \mathcal{R}_{\Vec{n}}^{\dagger}(\phi) \, \mathcal{T}_{\Vec{k}}^{\dagger}(\mu_2) \, (- S_{\Vec{m}}) \, \mathcal{T}_{\Vec{k}}(\mu_2) \, \mathcal{R}_{\Vec{n}}(\phi) \, \mathcal{T}_{\Vec{e}_z}(\mu_1) \, e^{-i \, \frac{\pi N}{2}} \, \vert N/2 \rangle_x \\
    &= - \langle S_{\Vec{m}} \rangle (\phi),
\end{align*}
i.e. the resulting signals are constantly zero.

\noindent Apart from this, there may exist less severe restrictions on the geometrical degrees of freedom of $\myvec{n}$, $\myvec{m}$ and $\myvec{k}$ to assure symmetry or anti-symmetry around $\langle S_{\Vec{m}} \rangle (\phi=0)$.

\section{Derivation of our figure of merit from the mean squared error} \label{app:deltaphi}

\noindent For anti-symmetric protocols, the mean squared error is given by 
\begin{equation*}
    \epsilon(\phi) = \sum_m \left[ \Hat{\phi}(m) - \phi \right]^2 p(m \vert \phi).
\end{equation*}
Evaluating this at $\phi = 0$ and using a linear phase estimator gives
\begin{align*}
    \epsilon(\phi) &\overset{\phi = 0}{=} (\Delta \phi)^2 = \sum_m \left[ \Hat{\phi}(m) - 0 \right]^2 p(m \vert \phi) = \sum_m \Hat{\phi}^2(m) p(m \vert \phi) \\
    &\, \, = \sum_m \left( \frac{m}{\partial_\phi S_{\myvec{m}}} \vert_{\phi = 0} \right)^2 p(m \vert \phi) = \frac{1}{(\partial_\phi S_{\myvec{m}})^2 \vert_{\phi = 0}} \sum_m m^2 \, p(m \vert \phi) \\
    &\, \, = \frac{\langle S_{\myvec{m}}^2 \rangle \vert_{\phi = 0}}{(\partial_\phi S_{\myvec{m}})^2 \vert_{\phi = 0}} = \frac{(\Delta S_{\myvec{m}})^2}{(\partial_\phi S_{\myvec{m}})^2} \Big\vert_{\phi = 0}
\end{align*}

\noindent For symmetric protocols, we know that
\begin{equation*}
    S_{\myvec{m}}(\phi) = S_{\myvec{m}}(-\phi), \quad \partial_\phi S_{\myvec{m}}(\phi) = - \, \partial_\phi S_{\myvec{m}}(- \phi).
\end{equation*}
Since symmetric protocols have an extremum at $\phi = 0$, we measure the signal at $\phi = \pm \varphi$:
\begin{enumerate}
\begin{minipage}{0.25\textwidth}
    \item $(\hat{\phi} + \varphi) = \frac{m_+}{\partial_\phi S_{\myvec{m}}} \big\vert_{\phi = \varphi}$
\end{minipage}
\begin{minipage}{0.35\textwidth}
    \item $(\hat{\phi} - \varphi) = \frac{m_-}{\partial_\phi S_{\myvec{m}}} \big\vert_{\phi = - \varphi}$
\end{minipage}
\end{enumerate}
A suitable estimator for the phase uncertainty is therefore
\begin{align*}
    \hat{\phi} = \frac{1}{2} \left( \frac{m_+}{\partial_\phi S_{\myvec{m}}} \big\vert_{\phi = \varphi} + \frac{m_-}{\partial_\phi S_{\myvec{m}}} \big\vert_{\phi = - \varphi} \right) = \frac{1}{2} \frac{1}{\partial_\phi S_{\myvec{m}}} \Big\vert_{\phi = \varphi} \left( m_+ - m_- \right).
\end{align*}
We assume that the conditional probabilities $p(m_+ \vert \phi + \varphi)$ and $p(m_- \vert \phi - \varphi)$ are independent, i.e.
\begin{align*}
    p(m_+, m_- \vert \phi) = p(m_+ \vert \phi + \varphi) \cdot p(m_- \vert \phi - \varphi).
\end{align*}
With this, we derive
\begin{align*}
    \epsilon(\phi) &\, \,= \sum_{m_+, m_-} p(m_+ \vert \phi + \varphi) \, p(m_- \vert \phi - \varphi) \left[ \frac{1}{2} \frac{1}{\partial_\phi S_{\myvec{m}} \vert_{\phi = \varphi}} (m_+ - m_-) - \phi \right]^2 \\[4pt]
    &\overset{\phi \, = \, 0}{=} \sum_{m_+, m_-} p(m_+ \vert \phi + \varphi) \, p(m_- \vert \phi - \varphi) \left[ \frac{1}{2} \frac{1}{\partial_\phi S_{\myvec{m}} \vert_{\phi = \varphi}} \right]^2 \, (m_+ - m_-)^2 \\[4pt]
    &\, \, = \frac{1}{4} \frac{1}{(\partial_\phi S_{\myvec{m}})^2 \vert_{\phi = \varphi}} \, \Big( \langle S_{\myvec{m}}^2 \rangle \vert_{\phi = \varphi} + \underbrace{\langle S_{\myvec{m}}^2 \rangle \vert_{\phi = - \varphi}}_{= \, \langle S_{\myvec{m}}^2 \rangle \vert_{\phi = \varphi}} - 2 \, \langle S_{\myvec{m}} \rangle \vert_{\phi = \varphi} \, \underbrace{\langle S_{\myvec{m}} \rangle \vert_{\phi = - \varphi}}_{= \, \langle S_{\myvec{m}} \rangle \vert_{\phi = \varphi}} \Big) \\[4pt]
    &\, \, = \frac{1}{4} \frac{1}{(\partial_\phi S_{\myvec{m}})^2 \vert_{\phi = \varphi}} \, 2 \, \Big( \langle S_{\myvec{m}}^2 \rangle - \langle S_{\myvec{m}} \rangle^2 \Big) \Big\vert_{\phi = \varphi} \\[4pt]
    &\, \, = \frac{1}{2} \frac{(\Delta S_{\myvec{m}})^2}{(\partial_\phi S_{\myvec{m}})^2} \Big\vert_{\phi = \varphi}.
\end{align*}

\end{document}

%% file: Refs.bbl
%